\documentclass[runningheads,a4paper]{llncs}

\usepackage{amsmath}
\usepackage{amsthm}
\usepackage{amssymb}

\usepackage{graphicx}
\usepackage{tikz}
\usepackage{hyperref}

\usepackage{url}
\urldef{\mailsa}\path|{thomas.gabor, marie.kiermeier, andreas.sedlmeier}@ifi.lmu.de|
\urldef{\mailsb}\path|{bernhard.kempter, cornel.klein, horst.sauer, reiner.schmid, jan.wieghardt}@siemens.com| 
\newcommand{\keywords}[1]{\par\addvspace\baselineskip
\noindent\keywordname\enspace\ignorespaces#1}

\newcommand{\scen}{scenario}
\newcommand{\scens}{scenarios}
\newcommand{\Scen}{Scenario}

\theoremstyle{definition}
\newtheorem{mydef}{Definition}
\newtheorem{mycon}{Concept}

\newcommand\copyrighttext{%
  \footnotesize This is a pre-print of an article published in \emph{Margaria T., Steffen B. (eds)}, Leveraging Applications of Formal Methods, Verification and Validation (ISoLA 2018), Lecture Notes in Computer Science, vol 11246, Springer, Cham, 2018. The final authenticated version is available online at DOI:  \href{http://doi.org/10.1007/978-3-030-03424-5_10}{http://doi.org/10.1007/978-3-030-03424-5\_10}}
\newcommand\copyrightnotice{%
\begin{tikzpicture}[remember picture,overlay]
\node[anchor=south,yshift=10pt,xshift=10pt] at (current page.south) {\fbox{\parbox{\textwidth}{\copyrighttext}}};
\end{tikzpicture}%
}

\begin{document}

\mainmatter  

\title{Adapting Quality Assurance\\to Adaptive Systems:\\The Scenario Coevolution Paradigm}

\titlerunning{Adapting Quality Assurance to Adaptive Systems}

%
%
\author{Thomas Gabor$^1$\and Marie Kiermeier$^1$\and Andreas Sedlmeier$^1$\and Bernhard Kempter$^2$\and Cornel Klein$^2$\and Horst Sauer$^2$\and Reiner Schmid$^2$\and Jan Wieghardt$^2$}
\authorrunning{Lecture Notes in Computer Science: Authors' Instructions}

\institute{$^1$LMU Munich, $^2$Siemens AG\\
\mailsa\\
\mailsb}

%
%

\toctitle{Adapting Quality Assurance to Adaptive Systems: The Scenario Coevolution Paradigm}
\tocauthor{Thomas Gabor et al.}
\maketitle

\copyrightnotice{}

\begin{abstract}
From formal and practical analysis, we identify new challenges that self-adaptive systems pose to the process of quality assurance. When tackling these, the effort spent on various tasks in the process of software engineering is naturally re-distributed. We claim that all steps related to testing need to become self-adaptive to match the capabilities of the self-adaptive system-under-test. Otherwise, the adaptive system's behavior might elude traditional variants of quality assurance. We thus propose the paradigm of scenario coevolution, which describes a pool of test cases and other constraints on system behavior that evolves in parallel to the (in part autonomous) development of behavior in the system-under-test. Scenario coevolution offers a simple structure for the organization of adaptive testing that allows for both human-controlled and autonomous intervention, supporting software engineering for adaptive systems on a procedural as well as technical level.

\end{abstract}

\section{Introduction}

Until recently, the discipline of software engineering has mainly tackled the process through which humans develop software systems. In the last few years, current break-throughs in the fields of artificial intelligence and machine learning have enabled new possibilities that have previously been considered infeasible or just too complex to tap into with ``manual'' coding: Complex image recognition, natural language processing, or decision making as it is used in complex games are prime examples. The resulting applications are pushing towards a broad audience of users. However, as of now, they are mostly focused on non-critical areas of use, at least when implemented without further human supervision. Software artifacts generated via machine learning are hard to analyze, causing a lack of trustworthiness for many important application areas.

We claim that in order to reinstate levels of trustworthiness comparable to well-known classical approaches, we need not essentially reproduce the principles of classical software test but need to develop a new approach towards software testing. We suggest to develop a system and its test suite in a competitive setting where each sub-system tries to outwit the other. We call this approach \emph{scenario coevolution} and attempt to show the necessity of such an approach. We hope that trust in that dynamic (or similar ones) can help to build a new process for quality assurance, even for hardly predictable systems.

Following a top-down approach to the issue, we start in Section~\ref{sec:formal} by introducing a formal framework for the description of systems. We augment it to also include the process of software and system development. Section~\ref{sec:related-work} provides a short overview on related work. From literature review and practical experience, we introduce four core concepts for the engineering of adaptive systems in Section~\ref{sec:concepts}. In order to integrate these with our formal framework, Section~\ref{sec:scenarios} contains an introduction of our notion of scenarios and their application to an incremental software testing process. In Section~\ref{sec:applications} we discuss which effect scenario coevolution has on a selection of practical software engineering tasks and how it helps implement the core concepts. Finally, Section~\ref{sec:conclusion} provides a short conclusion.


\section{Formal Framework}\label{sec:formal}

In this section we introduce a formal framework as a basis for our analysis. We first build upon the framework described in \cite{holzl2011towards} to define adaptive systems and then proceed to reason about the influence of their inherent structure on software architecture.

\subsection{Describing Adaptive Systems}

We roughly adopt the formal definitions of our vocabulary related to the description of systems from \cite{holzl2011towards}: We describe a system as an arbitrary relation over a set of variables.

\begin{mydef}[System \cite{holzl2011towards}]\label{def:system}
	Let $I$ be a (finite or infinite) set, and let $\mathcal{V} = (V_i)_{i \in I}$ be a family of sets. A \emph{system} of type $\mathcal{V}$ is a relation $S$ of type $\mathcal{V}$.
\end{mydef}

Given a System $S$, an element $s \in S$ is called the state of the system. For practical purposes, we usually want to discern various parts of a system's state space. For this reason, parts of the system relation of type $\mathcal{V}$ given by an index set $J \subseteq I$, i.e., $(V_j)_{j \in J}$, may be considered \emph{inputs} and other parts given by a different index set may be considered \emph{outputs} \cite{holzl2011towards}. Formally, this makes no difference to the system. Semantically, we usually compute the output parts of the system using the input parts.

We introduce two more designated sub-spaces of the system relation: \emph{situation} and \emph{behavior}. These notions correspond roughly to the intended meaning of inputs and outputs mentioned before. The situation is the part of the system state space that fully encapsulates all information the system has about its state. This may include parts that the system does have full control over (which we would consider counter-intuitive when using the notion of ``input''). The behavior encapsulates the parts of the system that can only be computed by applying the system relation. Likewise, this does \emph{not} imply that the system has full control over the values. Furthermore, a system may have an \emph{internal state}, which is parts of the state space that are neither included in the situation nor in the behavior. When we are not interested in the internal space, we can regard a system as a mapping from situations to behavior, written $S = X \stackrel{Z}{\leadsto} Y$ for situations $X$ and behaviors $Y$, where $Z$ is the internal state of the system $S$. Using these notions, we can more aptly define some properties on systems. 
%
%

Further following the line of thought presented in \cite{holzl2011towards}, we want to build systems out of other systems. At the core of software engineering, there is the principle of re-use of components, which we want to mirror in our formalism. 

\begin{mydef}[Composition]
	Let $S_1$ and $S_2$ be systems of types $\mathcal{V}_1 = (V_{1,i})_{i \in I_1}$ and $\mathcal{V}_2 = (V_{2,i})_{i \in I_2}$, respectively. Let $\mathcal{R}(\mathcal{V})$ be the domain of all relations over $\mathcal{V}$. A \emph{combination operator} $\otimes$ is a function such that $S_1 \otimes S_2 \in \mathcal{R}(\mathcal{V})$ for some family of sets $\mathcal{V}$ with $V_{1,1}, ..., V_{1,m}, V_{2,1}, ..., V_{2,n} \in \mathcal{V}$.\footnote{In \cite{holzl2011towards}, there is a more strict definition on how the combination operator needs to handle the designated inputs and outputs of its given systems. Here, we opt for a more general definition.} The application of a combination operator is called \emph{composition}. The arguments to a combination operator are called \emph{components}.
\end{mydef}

Composition is not only important to model software architecture within our formalism, but it also defines the formal framework for interaction: Two systems interact when they are combined using a combination operator $\otimes$ that ensures that the behavior of (at least) one system is recognized within the situation of (at least) another system.

\begin{mydef}[Interaction]
	Let $S = S_1 \otimes S_2$ be a composition of  type $\mathcal{V}$ of systems $S_1$ and $S_2$ of type $\mathcal{V}_1$ and $\mathcal{V}_2$, respectively, using a combination operator $\otimes$. If there exist a $V_1 \in \mathcal{V}_1$ and a $V_2 \in \mathcal{V}_2$ and a relation $R \in V_1 \times V_2$ so that for all states $s \in S$, $(proj(s, V_1), proj(s, V_2)) \in R$, then the components $S_1$ and $S_2$ interact with respect to $R$.
\end{mydef}

We can model an open system $S$ as a combination $S = C \otimes E$ of a core system $C$ and its environment $E$, both being modeled as systems again.

Hiding some of the complexity described in \cite{holzl2011towards}, we assume we have a logic $\mathfrak{L}$ in which we can express a system goal $\gamma$. We can always decide if $\gamma$ holds for a given system, in which case we write $S \models \gamma$ for $\gamma(S) = \top$. Based on \cite{holzl2011towards}, we can use this concept to define an adaptation domain:

\begin{mydef}[Adaptation Domain \cite{holzl2011towards}] \label{def:adaptation-domain}
	Let $S$ be a system. Let $\mathcal{E}$ be a set of environments that can be combined with $S$ using a combination operator $\otimes$. Let $\Gamma$ be a set of goals. An \emph{adaptation domain} $\mathcal{A}$ is a set $\mathcal{A} \subseteq \mathcal{E} \times \Gamma$. $S$ can adapt to $\mathcal{A}$, written $S \Vdash \mathcal{A}$ iff for all $(E, \gamma) \in \mathcal{A}$ it holds that $S \otimes E \models \gamma$. 
\end{mydef}

\begin{mydef}[Adaptation Space \cite{holzl2011towards}]\label{def:adaptation-space}
	Let $\mathcal{E}$ be a set of environments that can be combined with $S$ using a combination operator $\otimes$. Let $\Gamma$ be set of goals. An \emph{adaptation space} $\mathfrak{A}$ is a set $\mathfrak{A} \subseteq \mathfrak{P}(\mathcal{E}, \Gamma)$. 
\end{mydef}

We can now use the notion of an adaptation space to define a preorder on the adaptivity of any two systems.

\begin{mydef}[Adaptation \cite{holzl2011towards}]\label{def:adaptation}
Given two systems $S$ and $S'$, $S'$ is at least as adaptive as $S$, written $S \sqsubseteq S'$ iff for all adaptation spaces $\mathcal{A} \in \mathfrak{A}$ it holds that $S \Vdash \mathcal{A} \Longrightarrow S' \Vdash \mathcal{A}$.
\end{mydef}

Both Definitions~\ref{def:adaptation-domain} and~\ref{def:adaptation-space} can be augmented to include soft constraints or optimization goals. This means that in addition to checking against boolean goal satisfaction, we can also assign each system $S$ interacting with an environment $E$ a \emph{fitness} $\phi(S \otimes E) \in F$, where $F$ is the type of fitness values. We assume that there exists a preorder $\preceq$ on $F$, which we can use to compare two fitness values. We can then generalize Definition~\ref{def:adaptation-domain} and~\ref{def:adaptation-space} to respect these optimization goals.

\begin{mydef}[Adaptation Domain for Optimization] \label{def:adaptation-domain-opt}
	Let $S$ be a system. Let $\mathcal{E}$ be a set of environments that can be combined with $S$ using a combination operator $\otimes$. Let $\Gamma$ be a set of Boolean goals. Let $F$ be a set of fitness values and $\preceq$ be a preorder on $F$. Let $\Phi$ be a a set of fitness functions with codomain $F$. An \emph{adaptation domain} $\mathcal{A}$ is a set $\mathcal{A} \subseteq \mathcal{E} \times \Gamma \times \Phi$. $S$ can adapt to $\mathcal{A}$, written $S \Vdash \mathcal{A}$ iff for all $(E, \gamma, \phi) \in \mathcal{A}$ it holds that $S \otimes E \models \gamma$.
\end{mydef}

Note that in Definition~\ref{def:adaptation-domain-opt} we only augmented the data structure for adaptation domains but did not actually alter the condition to check for the fulfillment of an adaptation domain. This means that for an adaptation domain $\mathcal{A}$, a system needs to fulfill all goals in $\mathcal{A}$ but is not actually tested on the fitness defined by $\phi$. We could define a fitness threshold $f$ we require a system $S$ to surpass in order to adapt to $\mathcal{A}$ in the formalism. But such a check, written $f \preceq \phi(S \otimes E)$, could already be included in the Boolean goals if we use a logic that is expressive enough.

Instead, we want to use the fitness function as soft constraints: We expect the system to perform as well as possible on this metric, but we do not (always) require a minimum level of performance. However, we can use fitness to define a fitness preorder on systems:

\begin{mydef}[Optimization]\label{def:optimization}
	Given two systems $S$ and $S'$ as well as an adaptation space $\mathcal{A}$, $S'$ is at least as optimal as $S$, written $S \preceq_\mathcal{A} S'$, iff for all $(E, \gamma, \phi) \in \mathcal{A}$ it holds that $\phi(S \otimes E) \preceq \phi(S' \otimes E)$.
\end{mydef}

\begin{mydef}[Adaptation with Optimization]\label{def:adaptation-opt}
Given two systems $S$ and $S'$, $S'$ is at least as adaptive as $S$ with respect to optimization, written $S \sqsubseteq^* S'$ iff for all adaptation domains $\mathcal{A} \in \mathfrak{A}$ it holds that $S \Vdash \mathcal{A} \Longrightarrow S' \Vdash \mathcal{A}$ and $S \preceq_\mathcal{A} S'$.
\end{mydef}

Note that so far our notions of adaptivity and optimization are purely extensional, which originates from the black box perspective on adaptation assumed in \cite{holzl2011towards}. 

\subsection{Constructing Adaptive Systems}

We now shift the focus of our analysis a bit away from the question ``When is a system adaptive?'' towards the question ``How is a system adaptive?''.  This refers to both questions of software architecture (i.e., which components should we use to make an adaptive system?) and questions of software engineering (i.e., which development processes should we use to develop an adaptive system?). We will see that with the increasing usage of methods of artificial intelligence, design-time engineering and run-time adaptation increasingly overlap \cite{wirsing2015software}.

\begin{mydef}[Adaptation Sequence]\label{def:adaptation-sequence}
	A series of $|I|$ systems $\mathcal{S} = (S_i)_{i\in I}$ with index set $I$ with a preorder $\leq$ on the elements of $I$ is called an \emph{adaptation sequence} iff for all $i, j \in I$ it holds that $i \leq j \Longrightarrow S_i \sqsubseteq^* S_j$
\end{mydef}

Note that we used adaptation with optimization in Definition~\ref{def:adaptation-sequence} so that a sequence of systems $(S_i)_{i\in I}$ that each fulfill the same hard constraints ($\gamma$ within a singleton adaptation space $\mathfrak{A} = \{\{(E, \gamma, \phi)\}\}$) can form an adaptation sequence iff for all $i, j \in I$ it holds that $i \leq j \Longrightarrow \phi(S_i \otimes E) \preceq \phi(S_j \otimes E)$. This is the purest formulation of an optimization process within our formal framework.\footnote{Strictly speaking, an optimization \emph{process} would further assume there exists an optimization relation $o$ from systems to systems so that for all $i, j \in I$ it holds that $i \leq j \Longrightarrow o(S_i, S_j)$. But for simplicity, we consider the sequence of outputs of the optimization process a sufficient representation of the whole process.}

Such an adaptation sequence can be generated by continuously improving a starting system $S_0$ and adding each improvement to the sequence. Such a task can both be performed by a team of human developers or standard optimization algorithms as they are used in artificial intelligence. Only in the latter case, we want to consider that improvement happening within our system boundaries. Unlike the previously performed black-box analysis of systems, the presence of an optimization algorithm within the system itself does have implications for the system's internal structure. We will thus switch to a more ``grey box'' analysis in the spirit of \cite{bruni2012conceptual}. 

\begin{mydef}[Self-Adaptation]\label{def:self-adaptation}
	A system $S_0$ is called \emph{self-adaptive} iff the sequence $(S_i)_{i \in \mathbb{N}, i < n}$ for some $n \in \mathbb{N}$ with $S_i = S_0 \otimes S_{i-1}$ for $0 < i < n$ and some combination operator $\otimes$ is an adaptation sequence.
\end{mydef}

Note that we could define the property of self-adaptation more generally by again constructing an index set on the sequence $(S_i)$ instead of using $\mathbb{N}$, but chose not to do so to not further clutter the notation. For most practical purposes, the adaptation is going to happen in discrete time steps anyway. It is also important to be reminded that despite its notation, the combination operator $\otimes$ does not need to be symmetric and likely will not be in this case, because when constructing $S_0 \otimes S_{i-1}$ we usually want to pass the previous instance $S_{i-1}$ to the general optimization algorithm encoded in $S_0$.\footnote{Constructing a sequence $S_i := S_{i-1} \otimes S_{i-1}$ might be viable formulation as well, but is not further explored in this work.} Furthermore, it is important to note that the constant sequence $(S)_{i \in \mathbb{N}}$ is an adaptation sequence according to our previous definition and thus every system is self-adaptive with respect to a combination operator $X \otimes Y =_\text{def} X$. However, we can construct non-trivial adaptation sequence using partial orders $\sqsubset$ and $\prec$ instead of $\sqsubseteq$ and $\preceq$. As these can easily be constructed, we do not further discuss their definitions in this paper. In \cite{holzl2011towards} a corresponding definition was already introduced for $\sqsubset$.

The formulation of the adaptation sequence used to prove self-adaptivity naturally implies some kind of temporal structure. So basing said structure around $\mathbb{N}$ implies a very simple, linear and discrete model of time. More complex temporal evolution of systems is also already touched upon in \cite{holzl2011towards}. As noted, there may be several ways to define such a temporal structure on systems. We refer to related and future work for a more intricate discussion on this matter.

So, non-trivial self-adaptation does imply some structure for any self-adaptive system $S$ of type $\mathcal{V} = (V_i)_{i \in I}$: Mainly, there needs to be a subset of the type $\mathcal{V}' \subseteq \mathcal{V}$ that is used to encode the whole relation behind $S$ so that the already improved instances can sufficiently be passed on to the general adaptation mechanism.

For a general adaptation mechanism (as we previously assumed to be part of a system) to be able to improve a system's adaptivity, it needs to be able to access some representation of its goals and its fitness function. This provides a grey-box view of the system. We remember that we assumed we could split a system $S$ into situation $X$, internal state $Z$ and behavior $Y$, written $S = X \stackrel{Z}{\leadsto} Y$. If $S$ is self-adaptive, it can form a non-trivial adaptation sequence by improving on its goals or its fitness. In the former case, we can now assume that there exists some relation $G \subseteq X \cup Z$ so that $S \models \gamma \iff G \models \gamma$ for a fixed $\gamma$ in a singleton-space adaptation sequence. In the latter case, we can assume that there exists some relation $F \subseteq X \cup Z$ so that $\phi(S) = \phi(F)$ for a fixed $\phi$ in a singleton-space adaptation sequence.

Obviously, when we want to construct larger self-adaptive systems using self-adaptive components, the combination operator needs to be able to combine said sub-systems $G$ and/or $F$ as well. In the case where the components' goals and fitnesses match completely, the combination operator can just use the same sub-system twice. However, including the global goals or fitnesses within each local component of a system does not align with common principles in software architecture (such as encapsulation) and does not seem to be practical for large or open systems (where no process may ensure such a unification). Thus, constructing a component-based self-adaptive system requires a combination operator that can handle potentially conflicting goals and fitnesses. We again define such a system for a singleton adaptation space $\mathfrak{A} = \{\{(E, \gamma, \phi)\}\}$ and leave the generalization to all adaptation spaces out of the scope of this paper. 

\begin{mydef}[Multi-Agent System]\label{def:mas}
	Given a system $S = S_1 \otimes ... \otimes S_n$ that adapts to $\mathcal{A} = \{(E, \gamma, \phi)\}$. Iff for each $1 \leq i \leq n$ with $i, n \in \mathbb{N}, n > 1$ there is an adaptation domain $\mathcal{A}_i = \{(E_i, \gamma_i, \phi_i)\}$ so that (1) $E_i = E \otimes S_1 \otimes ... \otimes S_{i-1} \otimes S_{i+1} \otimes ... \otimes S_n$ and (2) $\gamma_i \neq \gamma$ or $\phi_i \neq \phi$ and (3) $S_i$ adapts to $\mathcal{A}_i$, then $S$ is a \emph{multi-agent system} with agents $S_1, ..., S_n$.
\end{mydef}

For practical purposes, we usually want to use the notion of multi-agent systems in a transistive way, i.e., we can call a system a multi-agent system as soon as any part of it is a multi-agent system according to Definition\ref{def:mas}. Formally, $S$ is a multi-agent system if there are systems components $S', R$ so that $S = S' \otimes R$ and $S'$ is a multi-agent system. We argue that this transitivity is not only justified but a crucial point for systems development of adaptive systems: Agents tend to utilize their environment to fulfill their own goals and can thus ``leak'' their goals into other system components. Not that Condition (2) of Definition~\ref{def:mas} ensures that not every system constructed by composition is regarded a multi-agent system; it is necessary to feature agents with (at least slightly) differing adaptation properties.

For the remainder of this paper, we will apply Definition~\ref{def:mas} ``backwards'': Whenever we look at a self-adaptive system $S$, whose goals or fitnesses can be split into several sub-goals or sub-fitnesses we can regard $S$ as a multi-agent system. Using this knowledge, we can apply design patterns from multi-agent systems to all self-adaptive systems without loss of generality. Furthermore, we need to be aware that especially if we do not explicitly design multi-agent coordination between different sub-goals, such a coordination will be done implicitly. Essentially, there is no way around generalizing software engineering approaches for self-adaptive systems to potentially adversarial components.



\section{Related Work}\label{sec:related-work}

Many researchers and practitioners in recent years have already been concerned about the changes necessary to allow for solid and reliable software engineering processes for (self\nobreakdash-)adaptive systems. Central challenges were collected in \cite{salehie2009self}, where issues of quality assurance are already mentioned but the focus is more on bringing about complex adaptive behavior in the first place. The later research roadmap of \cite{de2013software} puts a strong focus on interaction patterns of already adaptive systems (both between each other and with human developers) and already dedicates a section to verification and validation issues, being close in mind to the perspective of this work. We fall in line with the roadmap further specified in \cite{bures2015software,belzner2016software,bures2017software}.

While this work largely builds upon \cite{holzl2011towards}, there have been other approaches to formalize the notion of adaptivity: \cite{oreizy1999architecture} discusses high-level architectural patterns that form multiple inter-connected adaptation loops. In \cite{arcaini2015modeling} such feedback loops are based on the MAPE-K model \cite{kephart2003vision}. While these approaches largely focus on the formal construction of adaptive systems, there have also been approaches that assume a (more human-centric or at least tool-centric) software engineering perspective \cite{elkhodary2010fusion,andersson2013software,gabor2016simulation,weyns2017software}. We want to discuss two of those on greater detail:

In the results of the \emph{ASCENS} (Autonomous Service Component ENSembles) project~\cite{wirsing2015software}, the interplay between human developers and autonomous adaptation has been formalized in a life-cycle model featuring separate states for each the development progress of each respective feedback cycle. Classical software development tasks and self-adaptation (as well as self-monitoring and self-awareness) are regarded as equally powerful contributing mechanisms for the production of software. Both can be employed in junction to steer the development process. In addition, ASCENS built upon a (in parts) similar formal notion of adaptivity \cite{bruni2012conceptual,nicola2014formal} and sketched a connection between adaptivity in complex distributed systems and multi-goal multi-agent learning \cite{holzl2015reasoning}.


\emph{ADELFE} (Atelier de D\'{e}veloppement de Logiciels \`{a} Fonctionnalit\'{e} Emergente) is a toolkit designed to augment current development processes to account for complex adaptive systems \cite{bernon2003tools,bernon2005engineering}. For this purpose, ADELFE is based on the Rational Unified Process (RUP) \cite{kruchten2004rational} and comes with tools for various tasks of software design. From a more scientific point of view, ADELFE is also based on the theory of adaptive multi-agent systems. For ADELFE, multi-agent systems are used to derive a set of stereotypes for components, which ease modeling for according types of systems. It thus imposes stronger restrictions on system design than our approach intends to.

Besides the field of software engineering, the field of artificial intelligence research is currently (re-)discovering a lot of the same issues the discipline of of engineering for complex adaptive systems faced: The highly complex and opaque nature of machine learning algorithms and the resulting data structures often forces black-box testing and makes possible guarantees weak. When online learning is employed, the algorithm's behavior is subject to great variance and testing usually needs to work online as well. The seminal paper \cite{amodei2016concrete} provides a good overview of the issues. When applying artificial intelligence to a large variety of products, rigorous engineering for this kind of software seems to be one of the major necessities lacking at the moment.

\section{Core Concepts of Future Software Engineering}\label{sec:concepts}
Literature makes it clear that one of the main issues of the development of self-adapting systems lies with \emph{trustworthiness}. Established models for checking systems (i.e., verification and validation) do not really fit the notion of a constantly changing system. However, these established models represent all the reason we have at the moment to trust the systems we developed. Allowing the system more degrees of freedom thus hinders the developers' ability to estimate the degree of maturity of the system they design, which poses a severe difficulty for the engineering progress, when the desired premises or the expected effects of classical engineering tasks on the system-under-development are hard to formulate.

To aid us control the development/adaptation progress of the system, we define a set of \emph{principles}, which are basically patterns for process models. They describe the changes to be made in the engineering process for complex, adaptive systems in relation to more classical models for software and systems engineering.

\begin{mycon}[System and Test Parallelism]\label{con:parallelism} The system and its test suite should develop in parallel from the start with controlled moments of interchange of information. Eventually, the test system is to be deployed alongside the main system so that even during runtime, on-going online tests are possible \cite{calinescu2012self}. This argument has been made for more classical systems as well and thus classical software test is, too, no longer restricted to a specific phase of software development. However, in the case of self-learning systems, it is important to focus on the evolution of test cases: The capabilities of the system might not grow as experienced test designers expect them to compared to systems entirely realized by human engineering effort. Thus, it is important to conceive and formalize how tests in various phases relate to each other.

\end{mycon}

\begin{mycon}[System vs. Test Antagonism]\label{con:antagonism} Any adaptive systems must be subject to an equally adaptive test. Overfitting is a known issue for many machine learning techniques. In software development for complex adaptive systems, it can happen on a larger scale: Any limited test suite (we expect our applications to be too complex to run a complete, exhaustive test) might induce certain unwanted biases. 
Ideally, once we know about the cases our system has a hard time with, we can train it specifically for these situations. For the so-hardened system the search mechanism that gave us the hard test cases needs to come up with even harder ones to still beat the system-under-test. Employing autonomous adaptation at this stage is expected to make that arms race more immediate and faster than it is usually achieved with human developers and testers alone.
\end{mycon}

\begin{mycon}[Automated Realization]\label{con:automated} Since the realization of tasks concerning adaptive components usually means the application of a standard machine learning process, a lot of the development effort regarding certain tasks tends to shift to an earlier phase in the process model. The most developer time when applying machine learning techniques, e.g.,  tends to be spent on gathering information about the problem to solve and the right setup of parameters to use;  the training of the learning agent then usually follows one of a few standard procedures and can run rather automatically.
However, preparing and testing the component's adaptive abilities might take a lot of effort, which might occur in the design and test phase instead of the deployment phase of the system life-cycle.
\end{mycon}

\begin{mycon}[Artifact Abstraction]\label{con:general} To provide room for and exploit the system's ability to self-adapt, many artifacts produced by the engineering process tend to become more general in nature, i.e., they tend to feature more open parameters or degrees of freedom in their description. In effect, in the place of single artifacts in a classical development process, we tend to find families of artifacts or processes generating artifacts when developing a complex adaptive system. As we assume that the previously only static artifact is still included in the set of artifacts available in its place now, we call this shift ``generalization'' of artifacts. Following this change, many of the activities performed during development shift their targets from concrete implementations to more general artifact, i.e., when building a test suite no longer yields a series of runnable test cases but instead produces a test case generator.
When this principle is broadly applied, the development activities shift towards ``meta development''. The developers are concerned with setting up a process able to find good solutions autonomously instead of finding the good solutions directly.
\end{mycon}

\section{Scenarios}\label{sec:scenarios}

We now want to include the issue of testing adaptive systems in our formal framework. We recognize that any development process for systems following the principles described in Section~\ref{sec:formal} produces two central types of artifacts: The first one is a system $S = X \stackrel{Z}{\leadsto} Y$ with a specific desired behavior $Y$ so that it manages to adapt to a given adaptation space. The second is a set of situations, test cases, constraints, and checked properties that this system's behavior has been validated against. We call artifacts of the second type by the group name of \emph{\scens}.

\begin{mydef}[Scenario]\label{def:scenario}
	Let $S = X \stackrel{Z}{\leadsto} Y$ be a system and $\mathcal{A} = \{(E, \gamma, \phi)\}$ a singleton adaptation domain. A tuple $c = (X, Y, g, f), g \in \{\top, \bot \}, f \in \text{cod}(\phi)$ with $g = \top \iff S \otimes E \models \gamma$ and $f = \phi(S \otimes E)$ is called \emph{scenario}.\footnote{If we are only interested in the system's performance and not \emph{how} it was achieved, we can redefine a scenario to leave out $Y$.}
\end{mydef}

Semantically, scenarios represent the experience gained about the system's behavior during development, including both successful ($S \vDash \gamma$) and unsuccessful ($S \nvDash \gamma$) test runs. As stated above, since we expect to operate in test spaces we cannot cover exhaustively, the knowledge about the areas we did cover is an important asset and likewise result of the systems engineering process.

Effectively, as we construct and evolve a system $S$ we want to construct and augment a set of scenarios $C = \{c_1, ..., c_n\}$ alongside with it. $C$ is also called a \emph{scenario suite} and can be seen as a toolbox to test $S$'s adaptation abilities with respect to a fixed adaptation domain $\mathcal{A}$.

While formally abiding to Definition~\ref{def:scenario}, scenarios can be encoded in various ways in practical software development, such as:

\paragraph{Sets of data points of expected or observed behavior.} Given a system $S' = X' \leadsto Y'$ whose behavior  is desirable (for example a trained predecessor of our system or a watchdog component), we can create scenarios $(X', Y', g', f')$ with $g' = \top \iff S' \otimes E_i \models \gamma_i$ and $f' = \phi_i(S' \otimes E_i)$ for an arbitrary amount of elements $(E_i, \gamma_i, \phi_i)$ of an adaptation domain $\mathcal{A} = \{(E_1, \gamma_1, \phi_1), ..., (E_n, \gamma_n, \phi_n)\}$.

\paragraph{Test cases the system mastered.} In some cases, adaptive systems may produce innovative behavior before we actively seek it out. In this cases, it is helpful to formalize the produced results once they have been found so that we can ensure that the system's gained abilities are not lost during further development or adaptation. Formally, this case matches the case for ``observed behavior'' described above. However, here the test case $(X, Y, g, f)$ already existed as a scenario, so we just need to update $g$ and $f$ (with the new and better values) and possibly $Y$ (if we want to fix the observed behavior).

\paragraph{Logical formulae and constraints.} Commonly, constraints can be directly expressed in the adaptation domain. Suppose we build a system against an adaptation domain $\mathcal{A} = \{(E_1, \gamma_1, \phi_1), ..., (E_n, \gamma_n, \phi_n)\}$. We can impose a hard constraint $\zeta$ on the system in this domain by constructing a constrained adaptation domain $\mathcal{A'} = \{(E_1, \gamma_1 \land \zeta, \phi_1), ..., (E_n, \gamma_n \land \zeta, \phi_n)\}$ given that the logic of $\gamma_1, ..., \gamma_n, \zeta$ meaningfully supports an operation like the logical ``and'' $\land$. Likewise a soft constraint $\psi$ can be imposed via $\mathcal{A'} = \{(E_1, \gamma_1, \max(\phi_1, \psi), ), ..., \allowbreak(E_n, \gamma_n, \max(\phi_n, \psi))\}$ given the definition of the operator $\max$ that trivially follows from using the relation $\preceq$ on fitness values. Scenarios $(X', Y', g', f')$ can then be generated against the new adaptation domain $\mathcal{A}$  by taking pre-existing scenarios $(X, Y, g, f)$ and setting $X' = X, Y' = Y, g = \top, f = \psi((X \leadsto Y) \otimes E)$.

\paragraph{Requirements and use case descriptions (including the system's degree of fulfilling them).} If properly formalized, a requirement or use case description contains all the information necessary to construct an adaptation domain and can thus be treated as the logical formulae in the paragraph above. However, use cases are in practical development more prone to be incomplete views on the adaptation domain. We thus may want to stress the point that we do not need to update all elements of an adaptation domain when applying a constraint, i.e., when including a use case. We can also just add the additional hard constraint $\zeta$ or soft constraint $\psi$ to some elements of $\mathcal{A}$.

\paragraph{Predictive models of system properties.} For the most general case, assume that we have a prediction function $p$ so that $p(X) \approx Y$, i.e., the function can roughly return the behavior $S = X \leadsto Y$ will or should show given $X$. We can thus construct the predicted system $S' = X \leadsto p(X)$ and construct a scenario $(X, p(X), g, f)$ with $g = \top \iff S' \otimes E \models \gamma$ and $f = \phi(S' \otimes E)$. 

\paragraph{}

All of these types of artifacts will be subsumed under the notion of {\scens}. We can use them to further train and improve the system and to estimate its likely behavior as well as to perform tests (and ultimately verification and validation activities). 

\emph{{\Scen} coevolution} describes the process of developing a set of scenarios to test a system during the system-under-tests's development. Consequently, it needs to be designed and controlled as carefully as the evolution of system behavior \cite{arcuri2007coevolving,fraser2013whole}.

\begin{mydef}[Scenario Hardening]
	Let $c_1 = (X_1, Y_1, g_1, f_1)$ and $c_2 = (X_2, Y_2,\allowbreak g_1, f_2)$ be scenarios for a system $S$ and an adaptation domain $\mathcal{A}$. Scenario $c_2$ is \emph{at least as hard} as $c_1$, written $c_1 \leq c_2$, iff $g_1 = \top \implies g_2 = \top$ and $f_1 \leq f_2$.
\end{mydef}

\begin{mydef}[Scenario Suite Order]
	Let $C = \{c_1, ..., c_m\}$ and $C' = \{c_1', ..., c_n'\}$ be sets of scenarios, also called scenarios suites. Scenario suite $C'$ is \emph{at least as hard} as $C$, written $C \sqsubseteq C'$, iff for all scenarios $c \in C$ there exists a scenario $c'\in C'$ so that $c \leq c'$.
\end{mydef}

\begin{mydef}[Scenario Sequence]\label{def:scenario-sequence}
	Let $\mathcal{S} = (S_i)_{i\in I}, I = \{1, ..., n\}$ be an adaptation sequence for a singleton adaptation space $\mathfrak{A} = \{\mathcal{A}\}$. A series of sets $\mathcal{C} = (C_i)_{i \in I}$ is called a scenario sequence iff for all $i \in I, i < n$ it holds that $C_i$ is a scenario suite for $S_i$ and $\mathcal{A}$ and $C_i \sqsubseteq C_{i+1}$.
\end{mydef}


 We expect each phase of development to further alter the set of {\scens} just as it does alter the system behavior. The {\scens} produced and used at a certain phase in development must match the current state of progress. Valid {\scens} from previous phases should be kept and checked against the further specialized system. When we do not delete any {\scens} entirely, the continued addition of {\scens} will ideally narrow down allowed system behavior to the desired possibilities. Eventually, we expect all activities of system test to be expressible as the generation or evaluation of scenarios. New scenarios may simply be thought up by system developers or be generated automatically.
%

Finding the right {\scens} to generate is another optimization problem to be solved during the development of any complex adaptive system. {\Scen} evolution represents a cross-cutting concern for all phases of system development. Treating {\scens} as first-class citizen among the artifacts produced by system development thus yields changes in tasks throughout the whole process model.

\section{Applications of Scenario Coevolution}\label{sec:applications}

Having both introduced a formal framework for adaptation and the testing of adaptive systems using scenarios, we show in this section how these frameworks can be applied to aid the trustworthiness of complex adaptive systems for practical use.

\subsection{Criticality Focus}

It is very important to start the scenario evolution process alongside the system evolution, so that at each stage there exists a set of scenarios available to test the system's functionality and degree of progress (see Concept~\ref{con:parallelism}). This approach mimics the concept of agile development where between each sprint there exists a fully functional (however incomplete) version of the system. The ceoncept of {\scen} evolution integrates seamlessly with agile process models.

In the early phases of development, the common artifacts of requirements engineering, i.e., formalized requirements, serve as the basis for the scenario evolution process. As long as the adaptation space $\mathfrak{A}$ remains constant (and with it the system goals),  system development should form an adaptation sequence. Consequently, scenario evolution should then form a scenario sequence for that adaptation sequence. This means (according to Definition~\ref{def:scenario-sequence}), the scenario suite is augmented with newly generated scenarios (for new system goals or just more specialized subgoals) or with scenarios with increased requirements on fitness.\footnote{Note that every change in $\mathfrak{A}$ starts new sequences.} Ideally, the scenario evolution process should lead the learning components on the right path towards the desired solution. The ability to re-assign fitness priorities allows for an arms race between adaptive system and scenario suite (see Concept~\ref{con:antagonism}).

\paragraph{Augmenting Requirements.} Beyond requirements engineering, it is necessary to include knowledge that will be generated during training and learning by the adaptive components. Mainly, recognized scenarios that work well with early version of the adaptive system should be used as checks and tests when the system becomes more complex. This approach imitates the optimization technique of importance sampling on a systems engineering level. There are two central issues that need to be answered in this early phase of the development process:

\begin{itemize}
    \item Behavior Observation: How can system behavior be generated in a realistic manner? Are the formal specifications powerful enough? Can we employ human-labeled experience?
    \item Behavior Assessment: How can the quality of observed behavior be adequately assessed? Can we define a model for the users' intent? Can we employ human-labeled review?
\end{itemize}

\paragraph{Breaking Down Requirements.} A central task of successful requirements engineering is to split up the use cases in atomic units that ideally describe singular features. In the dynamic world, we want to leave more room for adaptive system behavior. Thus, the requirements we formulate tend to be more general in notion. It is thus even more important to split them up in meaningful ways in order to derive new sets of scenarios. The following design axes (without any claim to completeness) may be found useful to break down requirements of adaptive systems:

\begin{itemize}
    \item Scope and Locality: Can the goal be applied/checked locally or does it involve multiple components? Which components fall into the scope of the goal? Is emergent system behavior desirable or considered harmful?
    \item Decomposition and Smoothness: Can internal (possibly more specific) requirements be developed? Can the overall goal be composed from a clear set of subgoals? Can the goal function be smoothened, for example by providing intermediate goals? Can subgoal decomposition change dynamically via adaptation or is it structurally static?
    \item Uncertainty and Interaction: Are all goals given with full certainty? Is it possible to reason about the relative importance of goal fulfillment for specific goals a priori? Which dynamic goals have an interface with human users or other systems?
\end{itemize}

\subsection{Adaptation Cooldown}

We call the problem domain available to us during system design the \emph{off-site domain}. It contains all {\scens} we think the system might end up in and may thus even contain contradicting {\scens}, for example. In all but the rarest cases, the situations one single instance of our system will face in its operating time will be just a fraction the size of the covered areas of the off-site domain. Nonetheless, it is also common for the system's real-world experience to include {\scens} not occurring in the off-site domain at all; this mainly happens when we were wrong about some detail in the real world. Thus, the implementation of an adaptation technique faces a problem not unlike the \emph{exploration/exploitation dilemma} \cite{vcrepinvsek2013exploration}, but on a larger scale: We need to decide, if we opt for a system fully adapted to the exact off-site domain or if we opt for a less specialized system that leaves more room for later adaptation at the customer's site. The point at which we stop adaptation happening on off-site {\scens} is called the off-site adaptation border and is a key artifact of the development process for adaptive systems.


In many cases, we may want the system we build to be able to evolve beyond the exact use cases we knew about during design time. The system thus needs to have components capable of \emph{run-time} or \emph{online adaptation}. In the wording of this work, we also talk about \emph{on-site adaptation} stressing that in this case we focus on adaptation processes that take place at the customer's location in a comparatively specific domain instead of the broader setting in a system development lab. Usually, we expect the training and optimization performed on-site (if any) to be not as drastic as training done during development. (Otherwise, we would probably have not specified our problem domain in an appropriate way.)
As the system becomes more efficient in its behavior, we want to gradually reduce the amount of change we allow. In the long run, adaptation should usually work at a level that prohibits sudden, unexpected changes but still manages to handle any changes in the environment within a certain margin. The recognized need for more drastic change should usually trigger human supervision first.

\begin{mydef}[Adaptation Space Sequence]\label{def:adaptation-sequence-spaces}
	Let $S$ be a system. A series of $|I|$ adaptation spaces $\mathbb{A} = (\mathfrak{A}_i)_{i\in I}$ with index set $I$ with a preorder $\leq$ on the elements of $I$ is called an \emph{adaptation domain sequence} iff for all $i, j \in I, i \leq j$ it holds that: $S$ adapts to $\mathfrak{A}_j$ implies that $S$ adapts to $\mathfrak{A}_i$. 
\end{mydef}


System development constructs an adaptation space sequence (c.f. Concept~\ref{con:general}), i.e., a sequence of increasingly specific adaptation domains. Each of those can be used to run an adaptation sequence (c.f. Definition~\ref{def:adaptation-sequence}) and a scenario sequence (c.f. Definition~\ref{def:scenario-sequence}, Concept~\ref{con:antagonism}) to test it.

For the gradual reduction of the allowed amount of adaptation for the system we use the metaphor of a ``cool-down'' process: The adaptation performed on-site should allow for less change than off-site adaptation. And the adaptation allowed during run-time should be less than what we allowed during deployment. This ensures that decisions that have once been deemed right by the developers are hard to change later by accident or by the autonomous adaptation process.

\subsection{Eternal Deployment}

For high trustworthiness, development of the test cases used for the final system test should be as decoupled from the on-going scenario evolution as possible, i.e., the data used in both processes should overlap as little as possible. Of course, following this guideline completely results in the duplication of a lot of processes and artifacts. Still, it is important to accurately keep track of the influences on the respective sets of {\scens}. A clear definition of the off-site adaptation border provides a starting point for when to branch off a {\scen} evolution process that is independent of possible {\scen}-specific adaptations on the system-under-test's side. Running multiple independent system tests (cf. ensemble methods \cite{dietterich2000ensemble,hart2017constructing}) is advisable as well. However, the space of available independently generated data is usually very limited.



For the deployment phase, it is thus of key importance to carry over as much information as possible about the genesis of the system we deploy into the run-time, where it can be used to look up the traces of observed decisions. The reason to do this now is that we usually expect the responsibility for the system to change at this point: Whereas previously, any system behavior was overseen by the developers who could potentially backtrack any phenomenon to all previous steps in the system development process, now we expect on-site maintenance to be able to handle any potential problem with the system in the real world, requiring more intricate preparation for maintenance tasks (c.f. Concept~\ref{con:automated}). We thus need to endow these new people with the ability to properly understand what the system does and why.

Our approach follows the vision of \emph{eternal system design} \cite{nierstrasz2008change}, which is a fundamental change in the way to treat deployment: We no longer ship a single artifact as the result of a complex development process, but we ship an image of the process itself (cf. Concept~\ref{con:general}). As a natural consequence, we can only ever add to an eternal system but hardly remove changes and any trace of them entirely. Using an adequate combination operator, this meta-design pattern is already implemented in the way we construct adaptation sequences (c.f. Definition~\ref{def:adaptation-sequence}): For example, given a system $S_i$ we could construct $S_{i+1} = X \stackrel{Z}{\leadsto} Y$ in a way so that $S_i$ is included in $S_{i+1}$'s internal state $Z$.

 As of now, however, the design of eternal systems still raises many unanswered questions in system design. We thus resort to the notion of {\scens} only as a sufficient system description to provide explanatory power at run-time and recommend to apply standard ``destructive updates'' to all other system artifacts.

\section{Conclusion}\label{sec:conclusion}

We have introduced a new formal model for adaptation and test processes using our notion of scenarios. We connected this model to concrete challenges and arising concepts in software engineering to show that our approach of scenario coevolution is fit to tackle (a first few) of the problems when doing quality assurance for complex adaptive systems.

As already noted throughout the text, a few challenges still persist. Perhaps most importantly, we require an adequate data structure both for the coding of systems and for the encoding of test suites and need to prove the practical feasibility of an optimization process governing the software development life-cycle. For performance reasons, we expect that some restrictions on the general formal framework will be necessary. In this work, we also deliberately left out the issue of meta-processes: The software development life-cycle can itself be regarded as system according to Definition~\ref{def:system}. While this may complicate things at first, we also see potential in not only developing a process of establishing quality and trustworthiness but also a generator for such processes (akin to Concept~\ref{con:general}).

Systems with a high degree of adaptivity and, among those, systems employing techniques of artificial intelligence and machine learning will become ubiquitous. If we want to trust them as we trust engineered systems today, the methods of quality assurance need to rise to the challenge: Quality assurance needs to adapt to adaptive systems!

\bibliography{references}

\begin{thebibliography}{10}

\bibitem{holzl2011towards}
H{\"o}lzl, M., Wirsing, M.:
\newblock Towards a system model for ensembles.
\newblock In: Formal Modeling: Actors, Open Systems, Biological Systems.
\newblock Springer (2011)  241--261

\bibitem{wirsing2015software}
Wirsing, M., H{\"o}lzl, M., Koch, N., Mayer, P.:
\newblock Software Engineering for Collective Autonomic Systems: The ASCENS
  Approach. Volume 8998.
\newblock Springer (2015)

\bibitem{bruni2012conceptual}
Bruni, R., Corradini, A., Gadducci, F., Lafuente, A.L., Vandin, A.:
\newblock A conceptual framework for adaptation.
\newblock In: International Conference on Fundamental Approaches to Software
  Engineering, Springer (2012)  240--254

\bibitem{salehie2009self}
Salehie, M., Tahvildari, L.:
\newblock Self-adaptive software: Landscape and research challenges.
\newblock ACM transactions on autonomous and adaptive systems (TAAS) (2009)

\bibitem{de2013software}
De~Lemos, R., Giese, H., M{\"u}ller, H.A., Shaw, M., Andersson, J., Litoiu, M.,
  Schmerl, B., Tamura, G., Villegas, N.M., Vogel, T.,  et~al.:
\newblock Software engineering for self-adaptive systems: A second research
  roadmap.
\newblock In: Software Engineering for Self-Adaptive Systems II.
\newblock Springer (2013)  1--32

\bibitem{bures2015software}
Bures, T., Weyns, D., Berger, C., Biffl, S., Daun, M., Gabor, T., Garlan, D.,
  Gerostathopoulos, I., Julien, C., Krikava, F.,  et~al.:
\newblock Software engineering for smart cyber-physical systems---{T}owards a
  research agenda: Report on the first international workshop on software
  engineering for smart {CPS}.
\newblock ACM SIGSOFT Software Engineering Notes \textbf{40}(6) (2015)  28--32

\bibitem{belzner2016software}
Belzner, L., Beck, M.T., Gabor, T., Roelle, H., Sauer, H.:
\newblock Software engineering for distributed autonomous real-time systems.
\newblock In: Proceedings of the 2nd International Workshop on Software
  Engineering for Smart Cyber-Physical Systems, ACM (2016)  54--57

\bibitem{bures2017software}
Bures, T., Weyns, D., Schmer, B., Tovar, E., Boden, E., Gabor, T.,
  Gerostathopoulos, I., Gupta, P., Kang, E., Knauss, A.,  et~al.:
\newblock Software engineering for smart cyber-physical systems: Challenges and
  promising solutions.
\newblock ACM SIGSOFT Software Engineering Notes \textbf{42}(2) (2017)  19--24

\bibitem{oreizy1999architecture}
Oreizy, P., Gorlick, M.M., Taylor, R.N., Heimhigner, D., Johnson, G.,
  Medvidovic, N., Quilici, A., Rosenblum, D.S., Wolf, A.L.:
\newblock An architecture-based approach to self-adaptive software.
\newblock IEEE Intelligent Systems and Their Applications \textbf{14}(3) (1999)
   54--62

\bibitem{arcaini2015modeling}
Arcaini, P., Riccobene, E., Scandurra, P.:
\newblock Modeling and analyzing {MAPE-K} feedback loops for self-adaptation.
\newblock In: Proceedings of the 10th International Symposium on Software
  Engineering for Adaptive and Self-Managing Systems, IEEE Press (2015)

\bibitem{kephart2003vision}
Kephart, J.O., Chess, D.M.:
\newblock The vision of autonomic computing.
\newblock Computer \textbf{36}(1) (2003)  41--50

\bibitem{elkhodary2010fusion}
Elkhodary, A., Esfahani, N., Malek, S.:
\newblock {FUSION}: a framework for engineering self-tuning self-adaptive
  software systems.
\newblock In: Proceedings of the 18th ACM SIGSOFT International Symposium on
  Foundations of Software Engineering, ACM (2010)

\bibitem{andersson2013software}
Andersson, J., Baresi, L., Bencomo, N., de~Lemos, R., Gorla, A., Inverardi, P.,
  Vogel, T.:
\newblock Software engineering processes for self-adaptive systems.
\newblock In: Software Engineering for Self-Adaptive Systems II.
\newblock Springer (2013)  51--75

\bibitem{gabor2016simulation}
Gabor, T., Belzner, L., Kiermeier, M., Beck, M.T., Neitz, A.:
\newblock A simulation-based architecture for smart cyber-physical systems.
\newblock In: Autonomic Computing (ICAC), 2016 IEEE International Conference
  on, IEEE (2016)  374--379

\bibitem{weyns2017software}
Weyns, D.:
\newblock Software engineering of self-adaptive systems: an organised tour and
  future challenges.
\newblock (2017)

\bibitem{nicola2014formal}
Nicola, R.D., Loreti, M., Pugliese, R., Tiezzi, F.:
\newblock A formal approach to autonomic systems programming: The {SCEL}
  language.
\newblock ACM Transactions on Autonomous and Adaptive Systems (TAAS)
  \textbf{9}(2) (2014) ~7

\bibitem{holzl2015reasoning}
H{\"o}lzl, M., Gabor, T.:
\newblock Reasoning and learning for awareness and adaptation.
\newblock In: Software Engineering for Collective Autonomic Systems.
\newblock Springer (2015)  249--290

\bibitem{bernon2003tools}
Bernon, C., Camps, V., Gleizes, M.P., Picard, G.:
\newblock Tools for self-organizing applications engineering.
\newblock In: International Workshop on Engineering Self-Organising
  Applications, Springer (2003)  283--298

\bibitem{bernon2005engineering}
Bernon, C., Camps, V., Gleizes, M.P., Picard, G.:
\newblock Engineering adaptive multi-agent systems: The {ADELFE} methodology.
\newblock In: Agent-Oriented Methodologies.
\newblock IGI Global (2005)  172--202

\bibitem{kruchten2004rational}
Kruchten, P.:
\newblock The rational unified process: an introduction.
\newblock Addison-Wesley Professional (2004)

\bibitem{amodei2016concrete}
Amodei, D., Olah, C., Steinhardt, J., Christiano, P., Schulman, J., Man{\'e},
  D.:
\newblock Concrete problems in {AI} safety.
\newblock arXiv preprint arXiv:1606.06565 (2016)

\bibitem{calinescu2012self}
Calinescu, R., Ghezzi, C., Kwiatkowska, M., Mirandola, R.:
\newblock Self-adaptive software needs quantitative verification at runtime.
\newblock Communications of the ACM \textbf{55}(9) (2012)  69--77

\bibitem{arcuri2007coevolving}
Arcuri, A., Yao, X.:
\newblock Coevolving programs and unit tests from their specification.
\newblock In: Proceedings of the 22nd IEEE/ACM International Conference on
  Automated Software Engineering, ACM (2007)  397--400

\bibitem{fraser2013whole}
Fraser, G., Arcuri, A.:
\newblock Whole test suite generation.
\newblock IEEE Transactions on Software Engineering \textbf{39}(2) (2013)
  276--291

\bibitem{vcrepinvsek2013exploration}
{\v{C}}repin{\v{s}}ek, M., Liu, S.H., Mernik, M.:
\newblock Exploration and exploitation in evolutionary algorithms: A survey.
\newblock ACM Computing Surveys (CSUR) \textbf{45}(3) (2013) ~35

\bibitem{dietterich2000ensemble}
Dietterich, T.G.,  et~al.:
\newblock Ensemble methods in machine learning.
\newblock Multiple classifier systems \textbf{1857} (2000)  1--15

\bibitem{hart2017constructing}
Hart, E., Sim, K.:
\newblock On constructing ensembles for combinatorial optimisation.
\newblock Evolutionary Computation (2017)  1--21

\bibitem{nierstrasz2008change}
Nierstrasz, O., Denker, M., G{\^\i}rba, T., Lienhard, A., R{\"o}thlisberger,
  D.:
\newblock Change-enabled software systems.
\newblock In: Software-Intensive Systems and New Computing Paradigms.
\newblock Springer (2008)  64--79

\end{thebibliography}
\bibliographystyle{splncs}

\end{document}